\documentclass{aa}

\usepackage{graphicx}
\usepackage{amsmath}
\usepackage{amssymb}
\usepackage{booktabs}
\usepackage{multirow}
\usepackage{siunitx}
\usepackage{natbib}
\usepackage{hyperref}

\usepackage{txfonts}
\newcommand{\oauroral}{[O {\scriptsize III}]$\lambda$4363 }
\newcommand{\otwo}{[O {\scriptsize II}]$\lambda\lambda$3727,3729 }
\newcommand{\othree}{[O {\scriptsize III}]$\lambda\lambda$4960,5008 }
\def\ts     {\thinspace} 

\def\oiii {\ifmmode{{\rm [O}{\rm \scriptstyle III}]}\else{[O\ts {\scriptsize III}]\ts}\fi}
\def\oii  {\ifmmode{{\rm [O}{\rm \scriptstyle II}]}\else{[O\ts {\scriptsize II}]\ts}\fi}
\def\ha   {\ifmmode{{\rm H}{\alpha}}\else{H$\alpha$\ts}\fi}
\def\hb   {\ifmmode{{\rm H}{\beta}}\else{H$\beta$\ts}\fi}
\def\hg  {\ifmmode{{\rm H}{\gamma}}\else{H$\gamma$\ts}\fi}
\usepackage{orcidlink}

\begin{document}

 \title{Galaxy protoclusters as drivers of cosmic reionization}
   \subtitle{II. Te-based metallicities of Lyman-$\alpha$ emitters}
   \titlerunning{Te-based metallicities of Lyman-$\alpha$ emitters}

   \author{C. Moya-Sierralta\orcidlink{0000-0002-8876-267X}
          \inst{1,3}   
        \and
    C. L. Martin
    \inst{2}
        \and
          L. F. Barrientos
    \inst{1}
    \and
    L. Infante
    \inst{1,3,4}
        \and
    J. González-López
    \inst{1,3}
        \and
    W. Hu
    \inst{5,6}
    \and
    A. L. Faisst
    \inst{7}
    \and
    {Y. Harikane}\orcidlink{0000-0002-6047-430X}
    \inst{8}
    \and
    A. M. Koekemoer\orcidlink{0000-0002-6610-2048}
    \inst{9}
    \and
    S. Malhotra
    \inst{10}
        \and
    M. Ouchi
    \inst{8}
    \and
    Z. Peng
    \inst{2}
    \and
    J. Rhoads
    \inst{10}
        \and
    J. Wang
    \inst{11,12}
        \and
        {J. R. Weaver}\thanks{Brinson Prize Fellow}\orcidlink{0000-0003-1614-196X}
        \inst{9}
    \and
    I. Wold
    \inst{10}
    \and
    J. Yang
    \inst{2}
    \and
    Z. Zheng
    \inst{13}
    }
   \institute{Instituto de Astrofísica, Pontificia Universidad Católica de Chile
              \\
              \email{cmoya@carnegiescience.edu}
              \and
            University of California Santa Barbara
            \and
            Las Campanas Observatory, Carnegie Science, Raúl Bitrán 1200, La Serena, Chile
            \and
            Instituto  de  Astrofísica de  la  Facultad  de  Ingeniería  y  Ciencias,  Universidad  Diego  Portales,  Av. Ejército Libertador 441, Santiago, Chile
            \and
            Department of Physics and Astronomy, Texas A\&M University, College Station, TX 77843-4242, USA
            \and
            George P. and Cynthia Woods Mitchell Institute for Fundamental Physics and Astronomy, Texas A\&M University, College Station, TX 77843-4242, USA
            \and
            Caltech/IPAC, 1200 E. California Blvd. Pasadena, CA 91125, USA
            \and
            Institute for Cosmic Ray Research, The University of Tokyo, 5-1-5 Kashiwanoha, Kashiwa, Chiba 277-8582, Japan
            \and
            Space Telescope Science Institute, 3700 San Martin Drive,
Baltimore, MD 21218, USA
            \and
            Astrophysics Science Division, NASA Goddard Space Flight Center, 8800 Greenbelt Road, Greenbelt, Maryland, 20771, USA
            \and
           Department of Astronomy, University of Science and Technology of China, Hefei, Anhui 230026, People's Republic of China
            \and
            School of Astronomy and Space Science, University of Science and Technology of China, Hefei 230026, People's Republic of China
            \and
            Shanghai Astronomical Observatory, Chinese Academy of Sciences, Shanghai 200030, People's Republic of China
            }
   \date{}

\abstract
  {Protoclusters in the epoch of reionization are believed to host a significant fraction of the total cosmic star formation rate density, which makes them crucial in determining the timing and topology of reionization. Understanding the physical properties of galaxies within them is thus essential. A key property to trace the evolutionary stage of a galaxy, the gas-phase metallicity, can now be directly measured in the early Universe thanks to JWST.}
  {We aim to measure the gas-phase metallicity of Lyman-$\alpha$ emitters (LAEs) in the LAGER-z7OD1 protocluster at $z \approx 6.93$ via the direct ($T_e$-based) method using the [O~{\scriptsize III}]$\lambda$4363 auroral line. By combining these metallicities with stellar masses, we aim to constrain the mass--metallicity relation (MZR) in dense environments and determine how the chemical buildup of protocluster LAEs compares to other early galaxy populations.}
  {We used deep JWST-NIRSpec/MSA G395H spectroscopy to measure the rest-frame optical emission lines of our sample. We derived metallicities using self-consistent analytic prescriptions, with the systematic uncertainty from dust extinction fully propagated into an asymmetric error budget via Monte Carlo simulations based on the observed Balmer decrements. Stellar masses were derived from BEAGLE spectral energy distribution fitting using JWST/NIRCam and UltraVISTA photometry. We also created a weighted spectral stack to study the high signal-to-noise mean properties of the sample.}
  {We obtain five secure detections of the [O~{\scriptsize III}]$\lambda$4363 line and one upper limit, yielding individual metallicities in the range of $12+\log(\rm O/H) \approx 7.2 - 8.0$. Our weighted spectral stack yields $12+\log(\rm O/H) = 7.76^{+0.06}_{-0.05}$, which is $\sim 0.26$ dex higher than stacks of LAEs at similar redshifts from JWST spectroscopic surveys. However, fitting the MZR yields a slope of $\gamma = 0.232 \pm 0.214$ and a normalization of $Z_0 = 7.888 \pm 0.199$, which demonstrates that our sources lie systematically below the MZR of lower-redshift galaxies and the recently reported protocluster relations at $z \sim 5-7$.}
  {The enhancement in absolute metallicity relative to general population LAEs suggests an accelerated early assembly driven by the overdensity. However, the offset below the MZR seen in recent high-redshift studies indicates a complex interplay between mass buildup and gas accretion. Because our selection strictly targets actively star-forming LAEs, we are probably capturing a violent phase of pristine gas accretion from the intergalactic medium. This massive influx fuels star formation but overpowers normal environmental metal retention, thereby systematically diluting the interstellar media of these specific protocluster members.} 

\maketitle

\section{Introduction}

Stars in galaxies process interstellar material, producing heavy elements that are then released into the interstellar medium (ISM) and increase its gas-phase metallicity. This enrichment is offset by the inflow of primordial gas from the intergalactic medium (IGM). The gas-phase metallicity of the ISM is therefore an indicator of a galaxy's evolutionary stage and history. Metallicities can be measured by observing the abundance of a given element with respect to hydrogen. A common approach is to measure the electron temperature and density of the medium using temperature- and density-sensitive emission lines (auroral and nebular lines). By solving the equilibrium equations for the level populations, one obtains the emissivities of these lines and can then determine the relative ionic abundances with respect to hydrogen \citep{osterbrock2006astrophysics}.

To trace this chemical enrichment, oxygen serves as the primary proxy for  ISM metal enrichment because it is the most abundant heavy element by mass and is synthesized rapidly on short timescales via alpha-capture processes in massive stars. Consequently, tracing oxygen abundances offers an instantaneous snapshot of a galaxy's star formation history. In the highly ionized, low-metallicity environments typical of the early Universe, oxygen exists predominantly in the doubly ionized state ($O^{++}$). Measuring the temperature of this high-ionization zone ($T_e(\mathrm{[OIII]})$) is therefore crucial to obtaining accurate total abundances. The primary tool to constrain this temperature is the rest-frame optical auroral line $\mathrm{[O\,III]}\lambda4363$. However, due to its high excitation potential and intrinsic faintness, the $\mathrm{[O\,III]}\lambda4363$ line has been historically difficult to detect in ground-based observations at high redshift, which renders direct-method metallicities at $z > 3$ nearly impossible to obtain with ground-based facilities.

With the advent of the James Webb Space Telescope (JWST), it is now possible to obtain the high-sensitivity, rest-frame optical spectra of galaxies in the epoch of reionization (EoR). Indeed, early JWST datasets—such as those from the Early Release Observations (ERO) and the Cosmic Evolution Early Release Science (CEERS) survey—revealed a prominent sample of high-redshift sources characterized by low metallicity and high ionization potential \citep{Schaerer2022-bk, Adams2023-ue, Carnall2023-vh, Ferrara2023-ao, Rhoads2023-jj, trump2023physical, Sanders2023-excitation}. Crucially, the detection of the faint $\mathrm{[O\,III]}\lambda4363$ auroral line in these early datasets allowed the robust measurement of metallicities via the direct $T_e$ method \citep{izotov2006chemical}. As more objects have been observed through deep spectroscopic programs like the JWST Advanced Deep Extragalactic Survey (JADES), it has become possible to build statistically significant samples of galaxies with secure auroral line detections in the early Universe \citep{laseter2024jades, morishita2024diverse},  with gravitational lensing further extending these direct-method samples down to the lowest stellar masses probed at these redshifts \citep{hsiao2026}.

The stellar mass complements this picture since it encodes the integrated history of star formation as gas is converted into stars over cosmic time. Its relationship with gas-phase metallicity, the mass-metallicity relation (MZR), is one of the most widely used diagnostics of galaxy assembly. The MZR does more than confirm that stellar nucleosynthesis has occurred: its normalization, slope, and scatter constrain how the baryon cycle operates, reflecting the ongoing balance between accretion of pristine or pre-enriched gas, star formation, and metal-loaded outflows. This same balance is captured more directly by the fundamental metallicity relation (FMR), which extends the MZR by introducing the star formation rate (SFR) as a third parameter: at fixed stellar mass, galaxies with higher SFRs tend to have lower metallicities, consistent with dilution of the ISM by accreting gas \citep{mannucci2010fundamental,andrews2013mass}. 

Observations from JWST have pushed MZR measurements out to $z > 6$ \citep{Curti2024,Chemerynska2024} and shown that galaxies can undergo substantial chemical enrichment within the first 500 Myr of cosmic history. Interpreting the relation at these epochs is not straightforward, since several distinct physical processes can leave similar imprints on its shape. Candidates include galaxy downsizing, in which massive systems enrich more rapidly than their lower-mass counterparts; a possible shift toward a top-heavy stellar initial mass function (IMF) under primordial conditions; and pre-enrichment of the circumgalactic medium (CGM) by feedback from early starbursts.

The large-scale environment adds a further complication. In clusters and protoclusters, the baryon cycle can be substantially perturbed, with mechanisms such as ram-pressure stripping, strangulation, and enhanced gas accretion reshaping the chemical scaling relations of member galaxies over time. These effects have been traced from subtle enrichment signatures in the local Universe \citep[e.g.,][]{ellison2009mass} to more disruptive gas dynamics at intermediate redshift \citep[e.g.,][]{Gupta_2016}. Measuring the MZR in primordial overdensities therefore offers a way to test whether environmental effects were already shaping the chemical evolution of galaxies in the early Universe.

Protoclusters, i.e., the precursors of massive galaxy clusters found at $z\sim0$, are of particular interest for current extragalactic studies. These massive structures are believed to host up to 50\% of the total cosmic SFR density during cosmic dawn, which makes them crucial in determining the timing and topology of reionization \citep{Chiang2017-tz}. Today, galaxy clusters are composed mainly of quiescent elliptical galaxies \citep{dressler1980galaxy,bamford2009galaxy}. Studying the evolutionary process of protoclusters from extended structures that host high star formation densities to more compact and quiescent systems is thus crucial for understanding both galaxy formation and the reionization process of the cosmos \citep{chiang2013ancient}.

The impact of environmental density on galaxy chemical evolution remains a subject of active debate. While some observations at high redshift suggest no significant environmental dependence on gas-phase metallicity \citep{Kacprzak_2015, Namiki_2019, Calabro22}, others have reported both metal enrichment \citep{kulas2013mass} and dilution \citep{Valentino_2015} in overdense regions. It remains unclear whether protocluster galaxies exhibit higher metal content than photometrically selected galaxies from the general population, particularly during the EoR, when environmental effects may be modulated by enhanced gas accretion or feedback.

A promising method to find protoclusters is via the Lyman-$\alpha$ emission line. Using custom narrow-band filters, it is possible to detect hundreds of sources emitting in Lyman-$\alpha$, including dense regions with a high concentration of Lyman-$\alpha$ emitters (LAEs). Given the ubiquity and intrinsic brightness of this emission line, it has been possible to detect protoclusters up to $z\sim 8$ \citep{Ouchi_2005,Wang_2005,castellano2016first,chanchaiworawit2017gran,castellano2018spectroscopic,Jiang_18,calvi2019mos,Harikane_2019,higuchi2019silverrush,rodriguez2020ionized,hu2021lyman}. Moreover, finding these Lyman-$\alpha$ protoclusters close to the EoR at $z\sim 7$ can also yield important insights about structure formation during this period.

Over the past four years, JWST has transformed both of these fronts. On the protocluster side, multi-object spectroscopy with the Near Infrared Spectrograph (NIRSpec) has made it possible to confirm galaxy overdensities directly through the simultaneous detection of rest-frame optical emission lines in several members at once, which has extended spectroscopically confirmed structures out to $z \sim 7.9-8.2$ \citep{morishita2023glass, helton2024overdensities}, well beyond the reach of photometric selection alone. On the emission-line side, deep NIRSpec spectroscopy has made rest-frame optical diagnostics routinely accessible in the EoR \citep{gardner2006james,gardner2023james}, which has enabled not just redshift confirmation but also the physical characterization of individual galaxies, including their metallicity, ionization state, and SFR. These two developments are beginning to converge, allowing astronomers to study the chemical state of massive structures \citep[e.g.,][]{umehata2025elpis}.

We aim to measure the direct-method gas-phase oxygen abundances of LAEs in the $z \approx 7$ protocluster LAGER-z7OD1. Our primary sample consists of six LAEs for which JWST/NIRSpec G395H spectroscopy provides coverage of both the [O III]$\lambda$4363 auroral line and at least one component of the [O III]$\lambda\lambda$4960,5008 doublet. These sources are a subset of the $z \sim 7$ overdensity first presented in \citet{hu2021lyman} and further characterized in \citet{martin2026galaxy} (hereafter Paper I). By focusing on these six targets, we can directly compare the chemical enrichment of protocluster members with findings from other studies of photometrically selected galaxies in the general population at similar redshifts, thereby constraining the impact of dense environments on early galaxy evolution.

The paper is structured as follows. In Section 2, we present our data, data reduction, and spectrum extraction. In Section 3, we present our methodology. Section 4 presents and discusses our results, and Section 5 summarizes our conclusions.

\section{Data}

We used observations from the JWST Near Infrared Camera (NIRCam) and NIRSpec of LAEs in the $z \approx 6.93$ protocluster LAGER-z7OD1 (program GO-01635, PI Martin). While eight LAEs were targeted in the western wing of the structure (as detailed in Paper I), this study focuses on the six sources with sufficient spectral coverage to derive direct-method metallicities. LAE-1 and LAE-14 were excluded from the metallicity analysis as their \oiii strong lines fall within the detector gap.

\subsection{Photometry}
Pre-imaging was obtained with JWST/NIRCam using the F150W2 and F444W filters, which provide coverage of the rest-frame UV and optical, respectively. We also used public Y-, J-, H-, and K-band imaging from the UltraVISTA survey, conducted with the Visible and Infrared Survey Telescope for Astronomy (VISTA). Aperture photometry was performed in a 2 arcsec aperture to ensure consistency between space- and ground-based data.
 
\subsection{Micro-shutter assembly observations}
We refer the reader to Paper I for more details on the observations and data reduction. Here we present a brief description of the data products.
Observations were conducted using the G395H grating and F290LP filter ($R \approx 2700$), covering the rest-frame range from \otwo to H$\alpha$ at $z \approx 6.93$. Out of the eight targeted sources, six have spectral coverage of both the \oauroral auroral line and at least one component of the \oiii doublet. Within this sample of six galaxies, we achieved secure auroral line detections (signal-to-noise ratio, S/N, $> 3$) for five objects, while LAE-8 is treated as an upper limit.

A master background was constructed from empty slits and subtracted for most targets. However, as documented in early JWST/NIRSpec spectroscopic surveys of faint high-redshift sources, the default master background subtraction in the \texttt{calwebb\_spec2} pipeline can frequently lead to severe over-subtraction of the sky background—resulting in unphysical negative continua and compromised line measurements due to time-varying backgrounds and residual 1/f noise \citep[e.g.,][]{trump2023physical, haro2023spectroscopic, Carnall2023-vh}. This over-subtraction was particularly pronounced for our faintest source, LAE-8. To address this, we implemented a customized local background subtraction for LAE-8, extracted directly from the spatial regions of its 2D spectrum free of source emission, which ensured robust and physically consistent line flux measurements.

\subsection{Error scaling and extraction}
To prevent the underestimation of uncertainties during the resampling step, we scaled the error arrays by factors of 1.6 to 1.8, following the methodology of \citet{haro2023spectroscopic}. These scaled uncertainties were propagated through all subsequent metallicity and mass calculations. 

\begin{figure}
    \centering
    \includegraphics[width=0.8\linewidth]{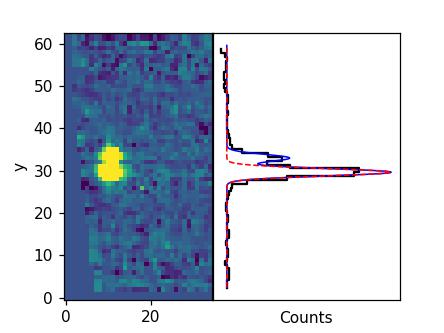}
    \caption{Example of our spatial profile fitting. Left: LAE-2 2D NIRSpec spectrum around the [O {\scriptsize III}]$\lambda$5008 line. Right: Spatial profile of the line emission. We fitted the data with a double Gaussian (solid blue). We then used the Gaussian model of the brightest component to extract the spectrum (dashed red).}
    \label{fig:spatial}
\end{figure}

Spatial profiles were measured from the brightest [O\,{\sc iii}] line and fitted with a Gaussian (or double Gaussian for multiple components). The 1D spectra were extracted by weighting along the spatial profile.  Figure \ref{fig:spatial} shows an example of how this was achieved.

\section{Methodology}

\subsection{The sample}
The parent sample consists of eight LAEs in the western part of the $z \approx 6.93$ protocluster LAGER-z7OD1, observed with JWST/NIRSpec as part of GO-01635 (Paper~I). Six of these sources have spectral coverage of the \oauroral line and at least one of the [O\,{\sc iii}]$\lambda\lambda$4960,5008 lines. Five sources show secure detections of \oauroral (S/N $> 3$), while LAE-8 provides only an upper limit. LAE-11 was retained despite large uncertainties as the auroral line is apparent in the 2D spectrum. Stellar masses for all targets were derived through spectral energy distribution (SED) fitting with the BayEsian Analysis of GaLaxy sEds tool (BEAGLE), using JWST/NIRCam + UltraVISTA photometry (details in Paper~I).

\subsection{Stellar masses and SED fitting}
To derive the stellar masses of our sample, we employed the BEAGLE tool \citep{Chevallard2016} following the exact configuration and multiband photometry measurements established in Paper I. We adopted their derived parameters for galaxy metallicity, ionization parameter, SFR, and color excess to define the shape of our spectral templates. 

For the BEAGLE modeling, the star formation history was assumed to be constant, with the SFR fixed and the duration of the current star formation episode treated as a free parameter. Crucially, the maximum stellar age was fixed to 700 Myr to reflect the age of the Universe at $z \sim 7$. We note that stellar mass estimates are sensitive to this maximum age assumption; for example, reducing this boundary to 10 Myr decreases the derived stellar masses by 0.6 to 0.8 dex for specific sources like LAE-8 and LAE-10, while the rest of the sample shows more moderate variations of $\pm0.1$ dex.

\subsection{Emission line fitting and aperture corrections}

\begin{figure*}
    \centering
    \includegraphics[width=0.49\linewidth]{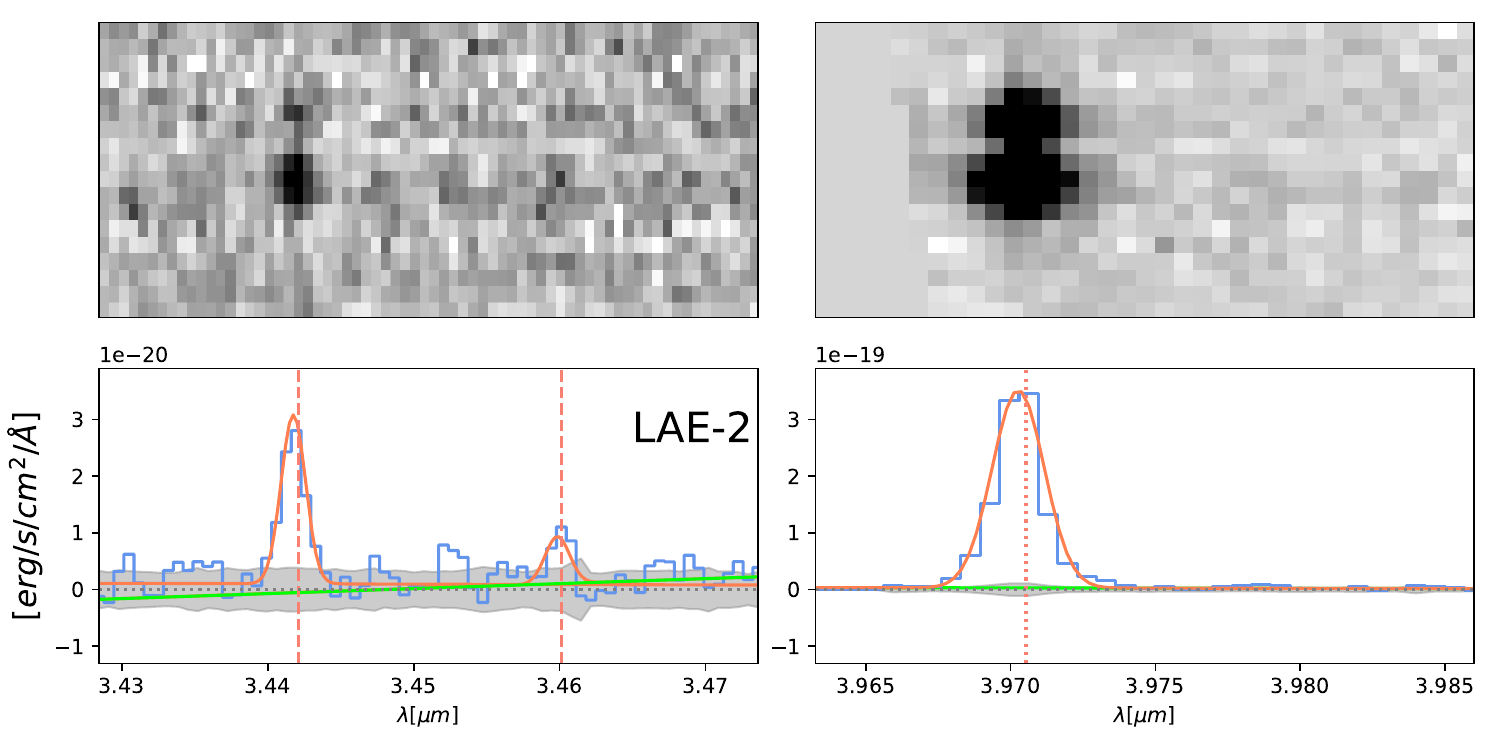}
    \includegraphics[width=0.49\linewidth]{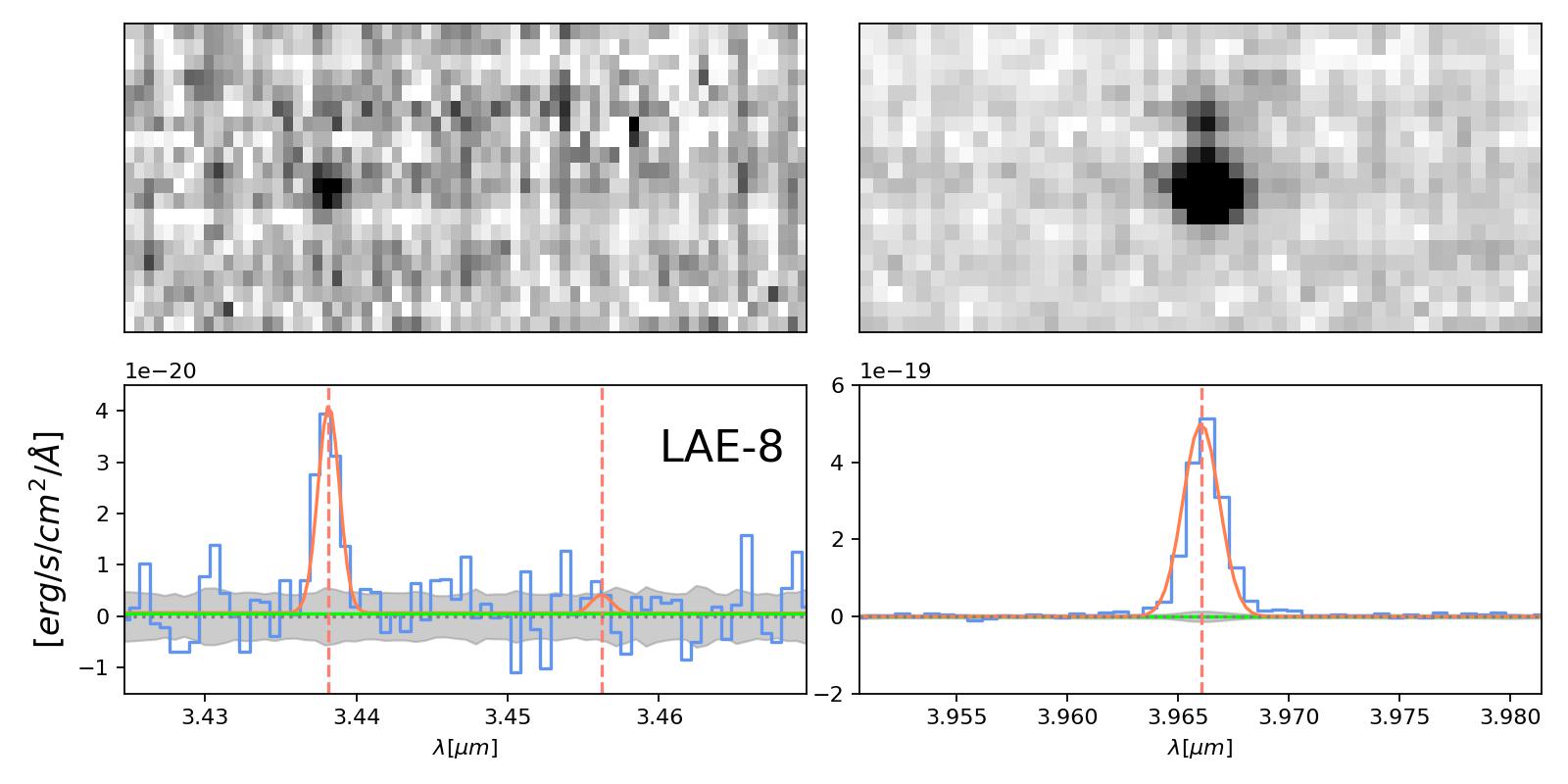}
    \includegraphics[width=0.49\linewidth]{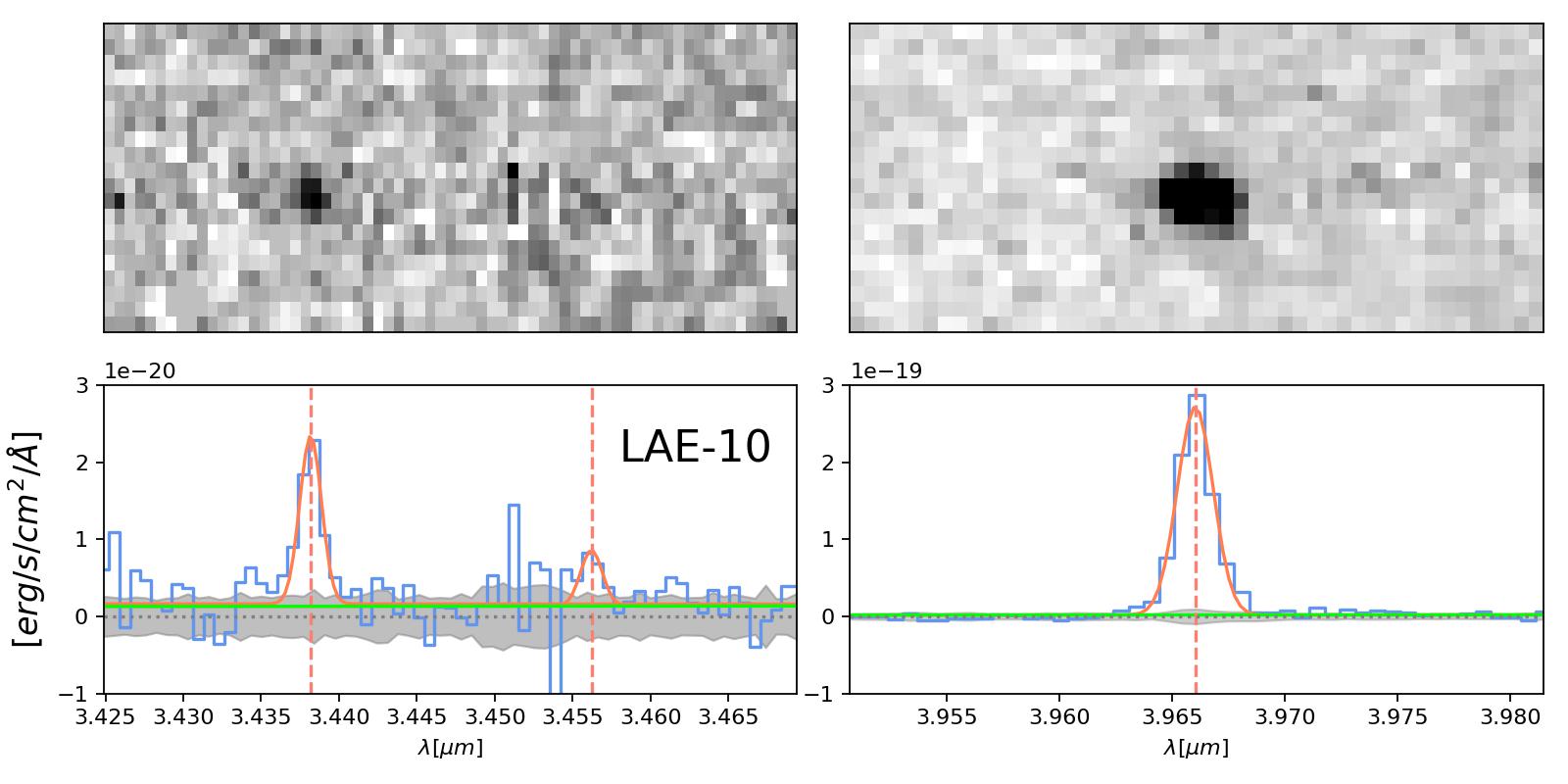}
    \includegraphics[width=0.49\linewidth]{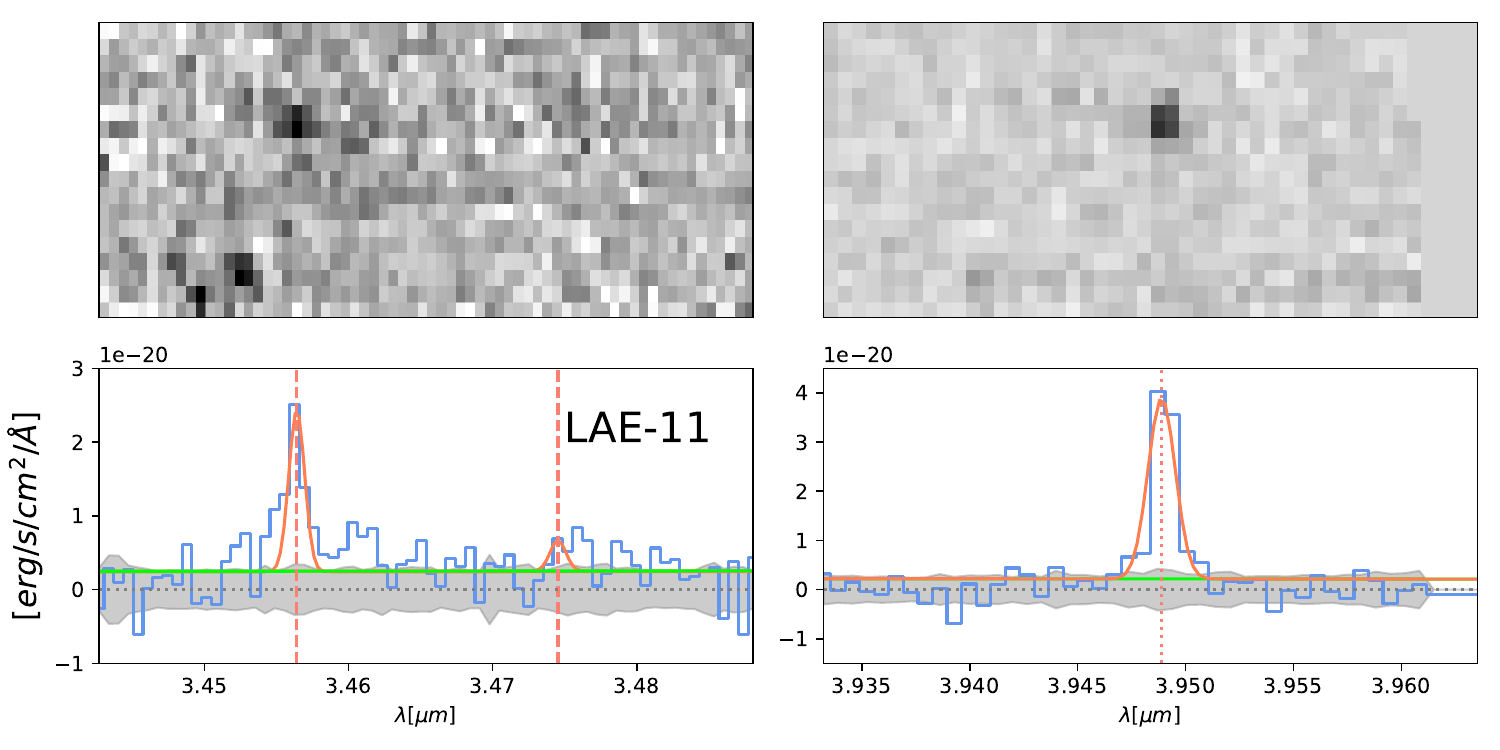}
    \includegraphics[width=0.49\linewidth]{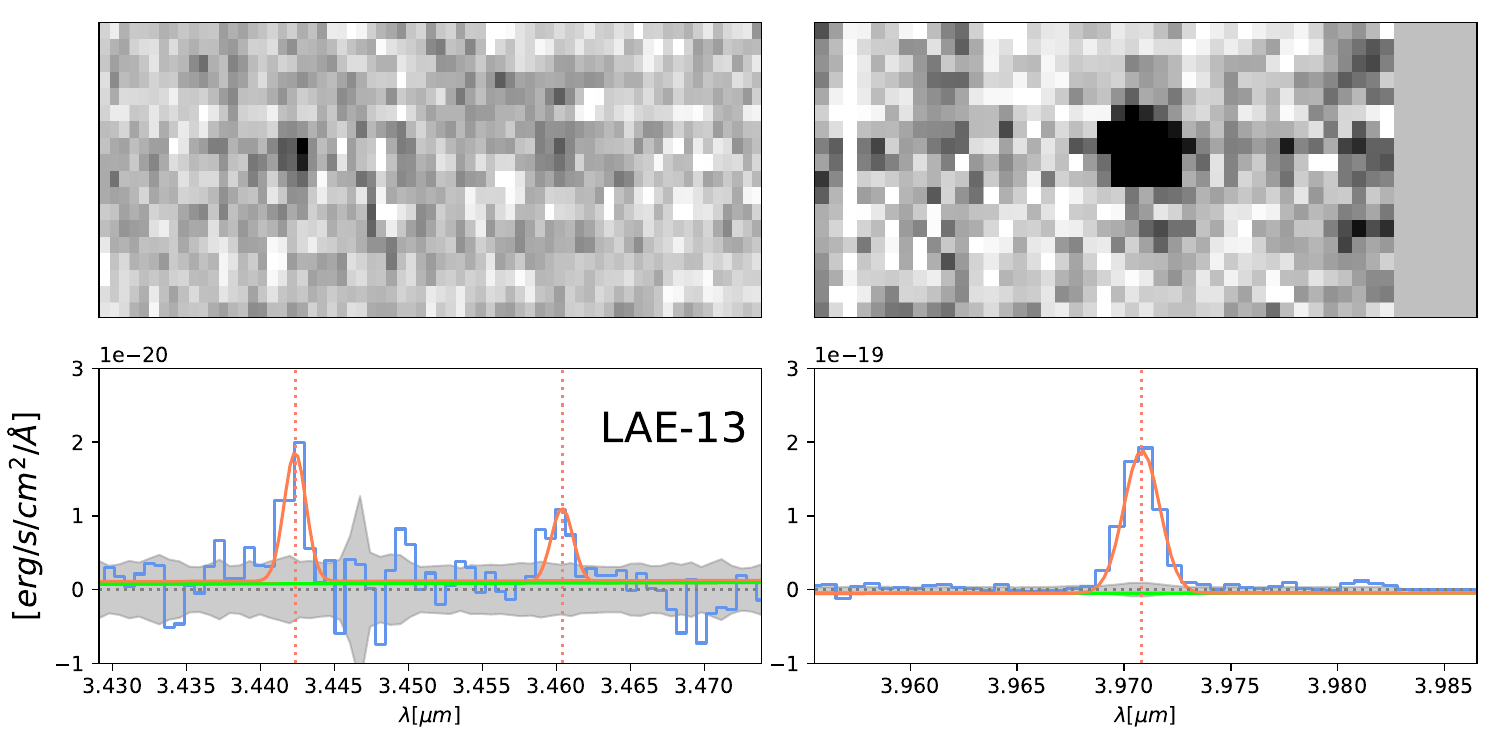}
    \includegraphics[width=0.49\linewidth]{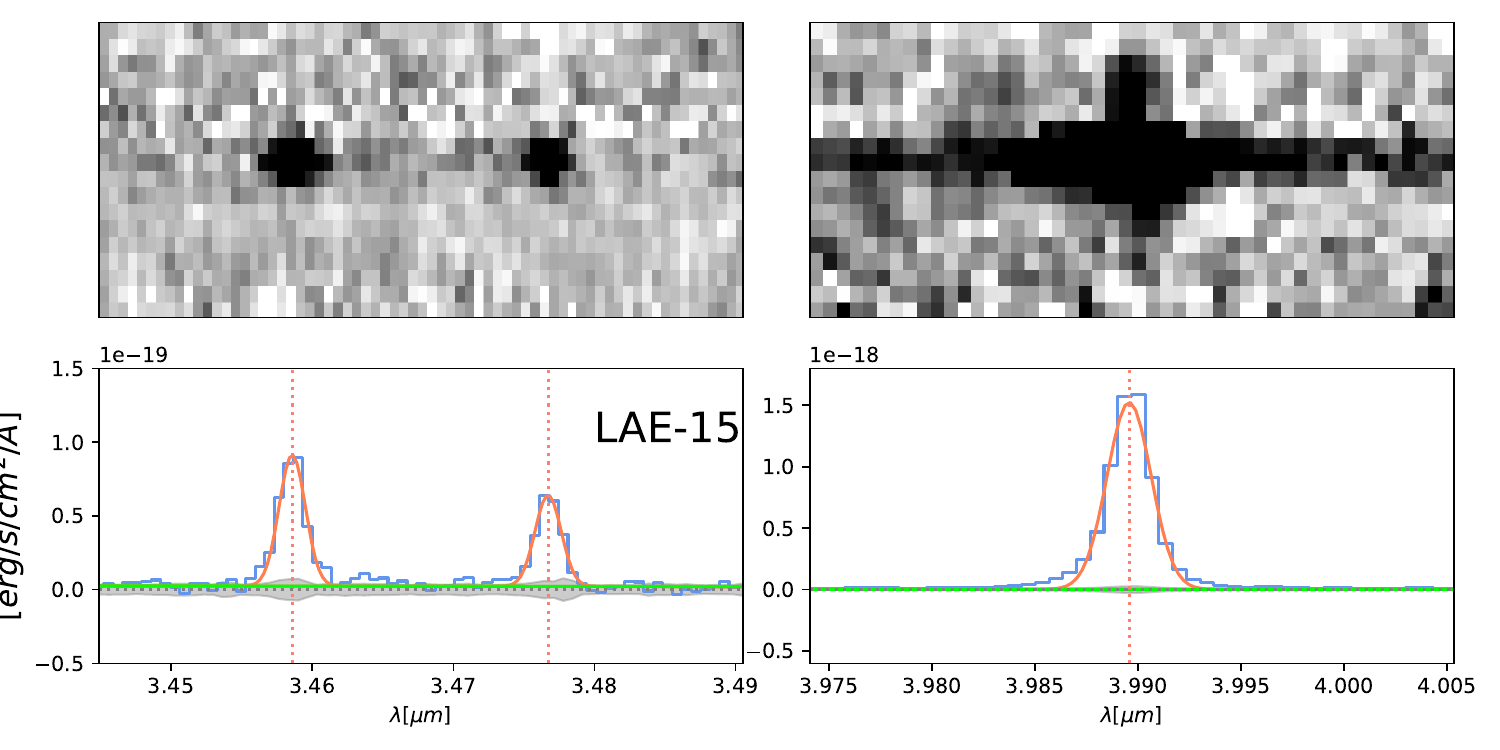}
    
    \caption{2D and 1D spectra of our sample. For every source, we plot the region around  H$\gamma$ and \oauroral, as well as the region around either [O\,{\sc iii}]$\lambda$5008 or [O\,{\sc iii}]$\lambda$4960. The data are shown in blue. Our Gaussian models are shown as orange lines. The noise is shown as a gray line. Dashed lines mark the position of the emission lines. Left column from top to bottom: LAE-2, LAE-10, LAE-13. Right column from top to bottom: LAE-8, LAE-11, LAE-15.}
    \label{fig:spec2d}
\end{figure*}

We simultaneously fitted our target emission lines and a continuum component using the Versatile Emission Line Fitting Tool \citep[\texttt{VerEmisFitting};][]{Peng_2025}. The width and systemic redshift were set to be the same for each line. While this might not be the case in principle, we found that setting free widths led to line fluxes consistent within 1$\sigma$ uncertainties, and setting free redshifts led to variations below $20\,\rm km\,s^{-1}$. Thus, fixing both parameters to be the same for each line was justified.
We fitted single- and double-Gaussian models to test whether a broad component was necessary. We found that the differences in derived metallicity were negligible; therefore, we adopted a single-component fit for the full sample. Figure \ref{fig:spec2d} shows the fits to relevant lines.

\begin{figure}
    \centering
    \includegraphics[width=0.8\linewidth]{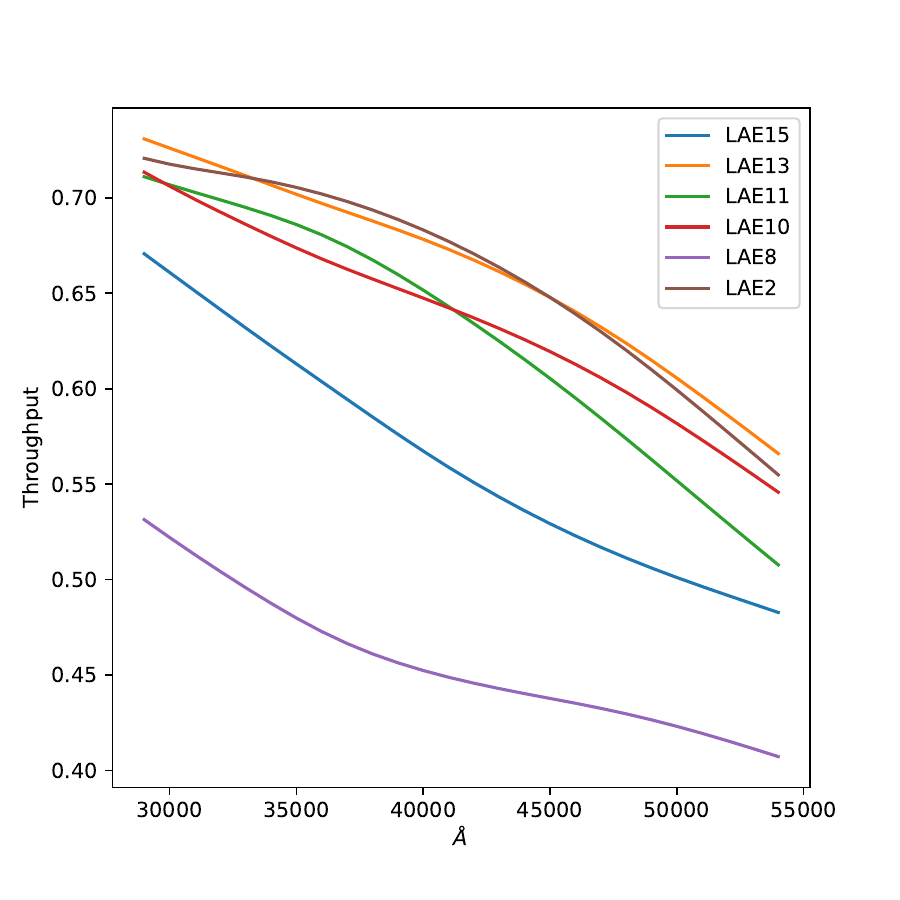}
    \caption{Throughput curves for each source in our sample. We note how the curves decrease toward a redder wavelength, which leads to a larger aperture correction of emission lines in that region than in the bluer end, such as \otwo.}
    \label{fig:aperture}
\end{figure}

To account for slit losses, we used the \texttt{MSAFIT} routine \citep{MSAFIT}.
The morphology of our sources was measured using \texttt{GALFIT} \citep{peng2011galfit}, and the resulting parameters were input to \texttt{MSAFIT}. We find that the aperture losses have small ($<10\%$) variations across the wavelength range. Nevertheless, aperture corrections slightly decrease the metallicity within the error bars. This is explained as the throughput curves show how the losses tend to increase toward redder wavelengths (see Fig. \ref{fig:aperture}). Thus, after applying the correction, the ratio between \otwo and \oiii decreases, leading to a lower contribution of singly ionized oxygen toward the total gas-phase metallicity.
More information on aperture corrections can be found in Appendix B of Paper I.

\subsection{Direct-method electron temperatures }
We first defined \oiii =[O~{\scriptsize III}]$\lambda$5008+[O~{\scriptsize III}]$\lambda$4960 for simplicity.
We aimed to measure the line ratio $\frac{\text{\oauroral}}{\text{\oiii}}$.  
When only one line of the \oiii  doublet is covered, we assumed [O~{\scriptsize III}]$\lambda$5008~=~2.98$\times$[O~{\scriptsize III}]$\lambda$4960 \citep{Storey2000-sp}. Following \cite{trump2023physical}, we took the ratio between the lines normalized by their neighboring Balmer emission lines:
\begin{equation}
    \frac{\text{\oauroral}}{\text{\othree}} = \frac{\text{\oauroral}}{H\gamma}\times(\frac{\hb}{\text{\othree}})\times\frac{1}{2.1}
.\end{equation}The 2.1 factor is the expected ratio between H${\beta}$ and \hg under case B recombination and corrects for dust attenuation and flux calibrations.  

We derived electron temperatures using the self-consistent analytic prescriptions of \citet{Perez-Montero2017-bm} to ensure methodological consistency.

The electron temperature of the high-ionization zone, $t([O III])$, was calculated as
\begin{equation}
t([O III]) = 0.7840 - 0.0001357 \times RO3 + \frac{48.44}{RO3}
\end{equation}
where $RO3 = I(4960 + 5008)/I(4363)$.

To measure the temperature of the low-ionization zone, $t_l$, we used the relation from \citet{Perez-Montero2017-bm}, which depends on both the high-ionization temperature and the electron density ($n_e$):
\begin{equation}
t_l = \frac{1.2 + 0.002\,n_e + 4.2/n_e}{t([O III])^{-1} + 0.08 + 0.003\,n_e + 2.5/n_e}
.\end{equation}
We fixed the electron density at $n_e = 300$ cm$^{-3}$ for all LAEs. This assumption is supported by the [O II]$\lambda\lambda 3729,3727$ doublet ratio measured from our high-S/N spectral stack, which yields $3729/3727 = 1.13 \pm 0.33$, consistent with this density regime \citep{osterbrock2006astrophysics}.

\subsection{Gas-phase metallicity and uncertainty propagation}
With the electron temperatures of the high- and low-ionization zones constrained, we determined the ionic abundances of singly and doubly ionized oxygen using the relations of \citet{Peng2023-ol}. Following a similar approach to \citet{Perez-Montero2017-bm}, these calibrations are obtained from fits to PyNeb \citep{Luridiana2015} using the default collision strengths of \oii and \oiii \citep{Kisielius2009, Storey2014}. We specifically adopted these calibrations because they offer precisions better than 0.01 dex within our expected temperature range ($T_e \approx 7000 - 25\,000$ K) and density regime ($n_e \approx 10 - 1000 \rm{cm}^{-3}$).

The relations for the ionic abundances are given by
 \citet{Peng2023-ol}:
\begin{equation}
\begin{aligned}
12 + \log \left(\frac{O^+}{H^+}\right) &= \log\left(\frac{\mathrm{[O\,II]}}{H\beta}\right) + 5.90 + \frac{1.65}{t_l} \\
&\quad - 0.54\,\log(t_l) + 7.99 \times 10^{-5} \, n_e
\end{aligned}
\end{equation}

\begin{equation}
\begin{split}
12 + \log \left(\frac{O^{++}}{H^+}\right) &= \log\left(\frac{\oiii}{H\beta}\right) + 6.21 + \frac{1.21}{t([O\,{\sc iii}])} \\
&\quad - 0.656\,\log\bigl(t([O\,{\sc iii}])\bigr)
\end{split}
.\end{equation}The total gas-phase oxygen abundance is the sum of these ionic ratios. To ensure a robust uncertainty budget, our final reported errors for $12+\log(O/H)$ include the uncertainties of both the \oii and \oiii components added in quadrature, as well as the intrinsic dispersion of the relations used.

\subsection{Interstellar dust extinction}
We investigated the impact of interstellar dust by comparing the observed H$\beta$/H$\gamma$ Balmer decrement for each source to the theoretical case B value of 2.1 (assuming $T_e = 2\times10^4$ K). While our individual galaxy measurements are consistent with this theoretical value within their respective error bars ($\approx 0.1$), we propagated the potential systematic uncertainty caused by dust into our final metallicity measurements.

Rather than applying a single extinction correction, we sampled the full range of extinction values allowed by the $1\sigma$ uncertainty on the measured Balmer decrement. Applying a \citet{calzetti2000dust} extinction curve, we used a Monte Carlo simulation (30\,000 realizations) to sample how these varying flux corrections impact the derived metallicities. Because the median shift from dust is small, we maintained the uncorrected line fluxes as our baseline values; however, the systematic uncertainty introduced by dust attenuation is fully integrated into the final asymmetric error bounds reported for our sample. Individual error bars for $E(B-V)$ can be found in Table 2 of Paper I.

\subsection{Spectral stacking}
\begin{figure}
    \centering
    \includegraphics[width=1\linewidth]{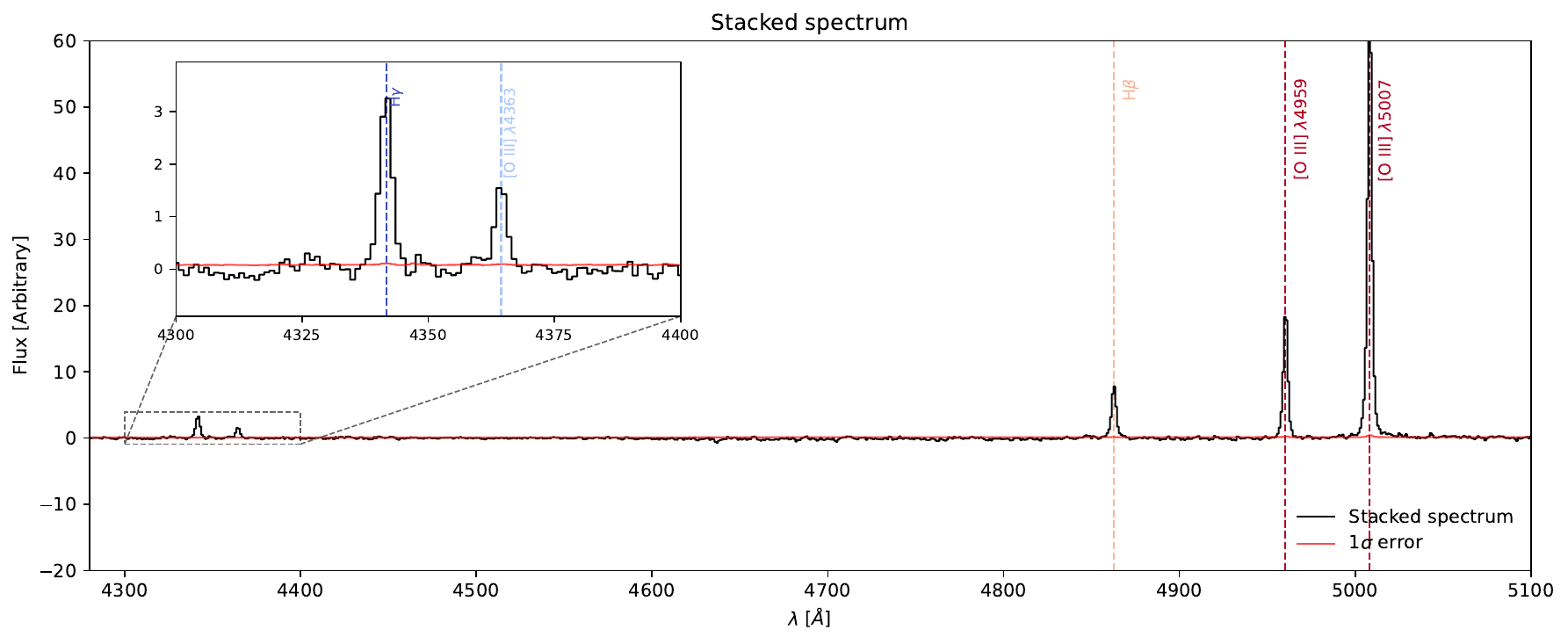}
    \caption{Spectral stack of five sources. LAE-11 was excluded given the lack of [O\,{\sc iii}]$\lambda$5008 coverage. The \hb and \oiii complex are clearly distinguished. The red line is the error array. An inset is shown displaying the \hg and \oauroral lines.}
    \label{fig:Stack}
\end{figure}

To compare our data to the spectral stacks presented by \cite{Hu2024} and \cite{Kumari2024}, we stacked the data of five out of six sources in our sample. LAE-2 and LAE-11 only cover [O~{\scriptsize III}]~$\lambda$5008 and [O~{\scriptsize III}]~$\lambda$4960 out of the \oiii doublet, respectively. Given that we require one of those lines to measure gas-phase metallicities, we stacked the sources containing [O~{\scriptsize III}]~$\lambda$5008, which has a higher S/N. Thus, we left LAE-11 out of the stacking process. Similarly to \cite{Hu2024} and \cite{Kumari2024}, we stacked the spectra weighting by their UV magnitude as measured in the F150W2 images. Each spectrum was corrected for aperture losses before stacking. The wavelength array was shifted to the rest frame using the systemic redshift of each source. The spectra were then rebinned onto an array covering 0.36--0.52\,$\mu$m with $10^{-4}\,\mu$m intervals. We note that after applying the redshift correction, the spectral pixel size of our sources is on the order of $8\times10^{-5}\,\mu$m.
The stacked spectrum is shown in Fig. \ref{fig:Stack}.
\section{Results and discussion} 

\begin{table*}[t]

\centering
  \caption{Emission line fluxes of the sample.}
\resizebox{\textwidth}{!}{
 \begin{tabular}{|c|c|c|c|c|c|c|c|}
 \hline
ID &  [O {\scriptsize II}]$\lambda$3726 &  [O {\scriptsize II}]$\lambda$3729 & \hg &  [O {\scriptsize III}]$\lambda$4363 & \hb &  [O {\scriptsize III}]$\lambda$4960 &  [O {\scriptsize III}]$\lambda$5008\\
\hline
LAE-2 & $28.78\pm5.58$ & $31.64\pm5.65$ & $40.53\pm4.40$ & $14.96\pm4.23$ & $101.91\pm6.57$ & X & $721.40\pm16.63$\\
LAE-8 & $18.45\pm10.83$ & $37.57\pm10.27$ & $72.15\pm8.73$ & $6.68\pm8.28$ & $132.62\pm10.33$ & $362.53\pm13.03$ & $1031.38\pm24.73$\\
LAE-10 & $19.55\pm7.06$ & $30.52\pm6.93$ & $34.54\pm5.05$ & $5.29\pm4.73$ & $80.26\pm6.63$ & $180.31\pm8.55$ &$561.72\pm17.39$\\
LAE-11 & $13.45\pm6.80$ & $4.76\pm6.01$ & $27.34\pm5.35$ & $5.80\pm4.62$ & $78.25\pm8.27$ & $65.88\pm7.48$ & X\\
LAE-13 & $19.51\pm6.92$ & $12.60\pm6.37$ & $27.07\pm5.14$ & $10.88\pm4.54$ & $73.14\pm5.34$ & $130.65\pm6.87$ &$433.62\pm14.35$\\
LAE-15 & $16.06\pm33.28$ & $15.73\pm33.75$ & $212.13\pm35.86$ & $140.17\pm30.71$ & $512.19\pm57.24$ & $1367.51\pm83.44$ & $4232.82\pm173.49$\\
\hline
\end{tabular}}
\tablefoot{Fluxes are given in units of $10^{-20}$ erg s$^{-1}$ cm$^{-2}$. The "X" represents a line not covered by the spectrum.}
   \label{tab:fluxes}
\end{table*}

\begin{table*}
\centering
\caption{Direct-method temperatures, metallicities, and stellar masses of the sample.}
\label{tab:results}
\begin{tabular}{lcccccc}
\hline\hline
ID & $\log\left(\frac{\rm [O\,III]\lambda4363}{\rm [O\,III]}\right)$ & $\log\left(\frac{\rm [O\,III]}{\rm H\beta}\right)$ & $T_e$ ($10^4$\,K) & $12+\log(\rm O/H)$ & $\log(M_*/M_\odot)$ \\
\hline
LAE-2  & $-1.85 \pm 0.13$ & $0.97 \pm 0.02$ & $1.52^{+0.26}_{-0.24}$ & $7.94^{+0.19}_{-0.17}$ & $8.93 \pm 0.02$ \\
LAE-8  & $<-2.38$         & $1.02 \pm 0.03$ & $< 1.10$               & $>8.05$                & $9.44 \pm 0.05$ \\
LAE-10 & $-1.80 \pm 0.14$ & $1.00 \pm 0.04$ & $1.45^{+0.32}_{-0.29}$ & $7.89^{+0.17}_{-0.16}$ & $9.42 \pm 0.02$ \\
LAE-11 & $-1.57 \pm 0.30$ & $0.53 \pm 0.05$ & $2.35^{+1.15}_{-0.95}$ & $7.22^{+0.45}_{-0.38}$ & $8.50 \pm 0.15$ \\
LAE-13 & $-1.43 \pm 0.12$ & $0.87 \pm 0.04$ & $2.58^{+0.71}_{-0.65}$ & $7.41^{+0.18}_{-0.17}$ & $8.67 \pm 0.32$ \\
LAE-15 & $-1.54 \pm 0.05$ & $1.06 \pm 0.02$ & $2.15^{+0.49}_{-0.44}$ & $7.61^{+0.17}_{-0.16}$ & $9.61 \pm 0.08$ \\
\hline
Stack  & $-1.65 \pm 0.04$ & $1.02 \pm 0.01$ & $1.89^{+0.12}_{-0.11}$ & $7.76^{+0.06}_{-0.05}$ & $9.14 \pm 0.12$ \\
\hline
\end{tabular}
\tablefoot{Results for our sample of five LAEs plus one upper limit. The reported gas-phase metallicity considers both singly and doubly ionized oxygen, derived using the \citet{Perez-Montero2017-bm} prescriptions. All errors come from propagating the statistical uncertainty in our temperature and metallicity measurements via Monte Carlo simulations.}
\end{table*}

\begin{figure}
    \centering
    \includegraphics[width=1\linewidth]{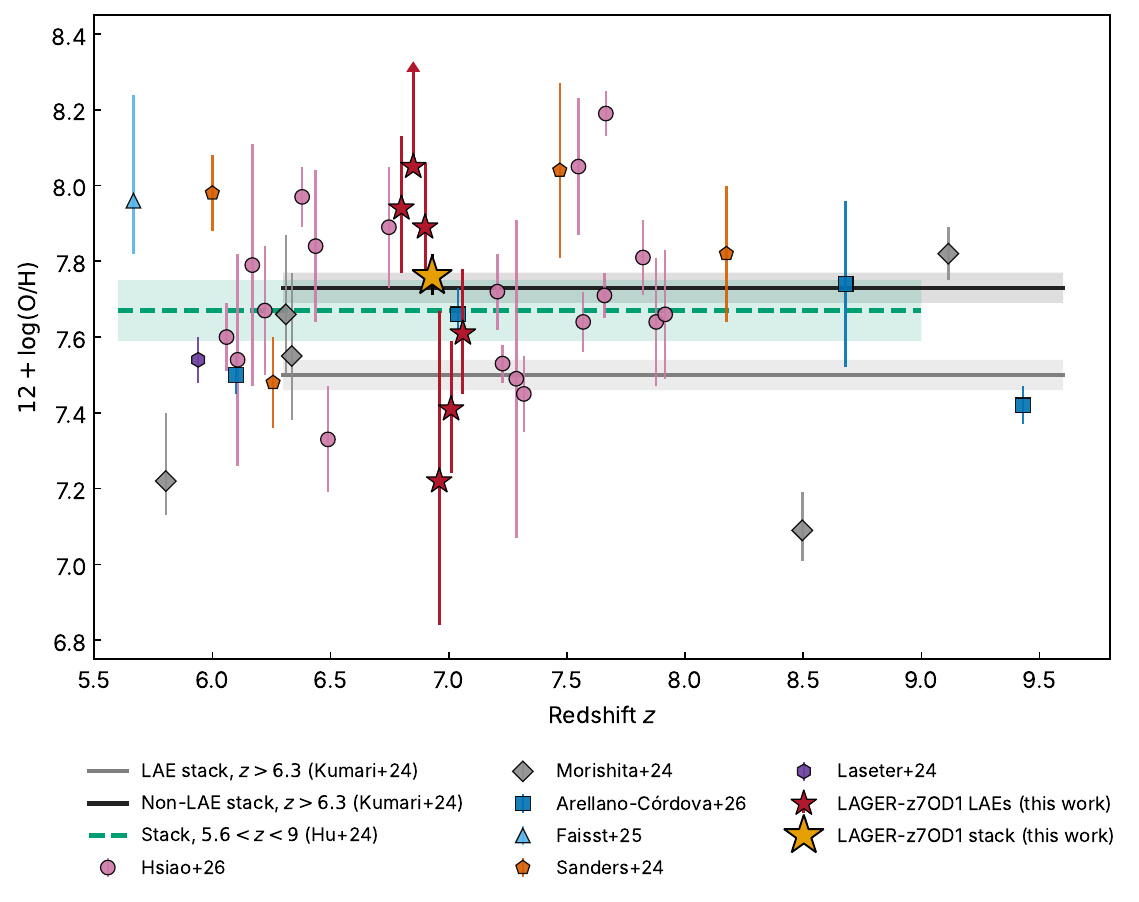}
    \caption{Direct-method metallicities as a function of redshift. Red stars show the individual LAGER-z7OD1 LAEs (Table~\ref{tab:results}), with LAE-8 shown as a lower limit, and the orange star shows our weighted stack. All our sources lie at $z\approx 6.93$; the individual points are spread along the x-axis for clarity. For comparison, we show individual galaxies at $z>5.5$ with [O\,{\sc iii}]$\lambda$4363-based metallicities from \citet{hsiao2026} (pink circles), \citet{morishita2024diverse} (gray diamonds), \citet{arellano2026} (blue squares), \citet{sanders2024direct} (orange pentagons), \citet{laseter2024jades} (purple hexagon), and \citet{faisst2025} (light blue triangle). Galaxies that appear in more than one of these samples are plotted once, using the most recent analysis. The gray and black solid lines mark the stacked metallicities of LAEs and non-LAEs at $z>6.3$ from \citet{Kumari2024}, and the green dashed line marks the stack of $5.6<z<9$ galaxies from \citet{Hu2024}; the shaded bands show their $1\sigma$ uncertainties over the redshift range of each stack. The metallicities within LAGER-z7OD1 are similar to those of other galaxies at $z>6$.}
    \label{fig:Z_vs_OH}
\end{figure}

We show our results in Table~\ref{tab:results}. We find that sources with \oauroral detections have low metallicities compared to local-Universe line emitters that have average gas-phase metallicities around 8 \citep{jiang2019direct}. However, their range agrees with that of high-$z$ objects observed with JWST (see Fig. \ref{fig:Z_vs_OH}). The mean metallicity for the five sources with secure auroral line detection is $12+\log(\rm O/H)=7.61\pm0.31$.

We present the gas-phase metallicities derived for our sample of LAEs using the direct ($T_e$-based) method. The measurements, obtained with the analytic prescriptions of \citet{Perez-Montero2017-bm}, are summarized in Table \ref{tab:results}. Five sources show secure detections of the [O III]$\lambda$4363 auroral line (S/N $> 3$), while LAE-8 yields only a lower limit on the metallicity. LAE-11 was retained in the analysis despite its large uncertainties. All values are calculated on the observed (uncorrected) line fluxes as our baseline; the median shift caused by the measured dust correction is included in the total error budget.

\subsection{Metallicities and environmental acceleration}
The five secure detections span the range $12 + \log(\rm O/H) = 7.2$ -- $8.0$, with a mean value of $7.61 \pm 0.31$. The weighted stack of five sources (excluding LAE-11) yields a higher S/N measurement of $12 + \log(\rm O/H) = 7.76^{+0.06}_{-0.05}$. 

When compared to continuum-selected LAEs from the general population at similar redshifts, our protocluster sources exhibit a distinct chemical enhancement. Specifically, our stacked metallicity of $7.76^{+0.06}_{-0.05}$ is approximately 0.26 dex higher than the stacked results of $12+\log(\rm O/H) = 7.50$ reported for general population LAEs at $z>6.3$ by \citet{Kumari2024}, and higher than the $7.67$ reported for $5.6 < z < 9$ galaxies by \citet{Hu2024}. 

This $\sim 0.25$ dex offset suggests that galaxies in the dense protocluster environment may have experienced accelerated star formation and chemical enrichment. Given that these LAEs lie within an overdensity, this enhanced growth could push them to higher masses compared to continuum-selected LAEs, and thus higher metallicities. We caution, however, that our error bars on the mean are large due to low number statistics, and LAEs generally trace the low-mass, low-metallicity regime; more massive sources with even higher metal contents might not be detected via Lyman-$\alpha$ emission. 

\subsection{The mass--metallicity relation (MZR)}

\begin{table*}
\centering
\caption{Best-fit parameters of the MZR.}
\label{tab:mzr_params}
\begin{tabular}{lcccc}
\hline\hline
Reference & Redshift & $Z_0$ & $\gamma$ & Note \\
\hline
This Work & $z \sim 6.9$ & $7.888 \pm 0.199$ & $0.232 \pm 0.214$ & Auroral line diagnostics \\
This Work & $z = 3$--$9$ & $8.054 \pm 0.026$ & $0.202 \pm 0.018$ & Combined fit\tablefootmark{a} \\
\citet{hsiao2026} & $z > 6$ & $8.205 \pm 0.214$ & $0.239 \pm 0.092$ & Fitted from sample \\
\citet{curti2020mass} & $z \sim 0$ & --- & --- & SDSS\tablefootmark{b} \\
\citet{Sanders2021} & $z \sim 2.3$ & $8.51 \pm 0.02$ & $0.30 \pm 0.02$ & MOSDEF \\
\citet{Sanders2021} & $z \sim 3.3$ & $8.41 \pm 0.03$ & $0.29 \pm 0.02$ & MOSDEF \\
\citet{li2023mass} & $z \sim 2$ & $8.18 \pm 0.02$\tablefootmark{c} & $0.13 \pm 0.04$ & Lensed dwarfs (stacks) \\
\citet{li2023mass} & $z \sim 3$ & $8.08 \pm 0.03$\tablefootmark{c} & $0.16 \pm 0.03$ & Lensed dwarfs (stacks) \\
\citet{nakajima2023jwst} & $z = 4$--$10$ & $8.24 \pm 0.05$ & $0.25 \pm 0.03$ & ERO, GLASS, CEERS \\
\citet{Curti2024} & $z = 3$--$10$ & $7.72 \pm 0.02$\tablefootmark{c} & $0.17 \pm 0.03$ & JADES \\
\citet{Curti2024} & $z = 6$--$10$ & $7.65 \pm 0.04$\tablefootmark{c} & $0.11 \pm 0.05$ & JADES \\
\citet{Li2026} & $z \sim 5$--$7$ & $8.34$\tablefootmark{d} & $0.33 \pm 0.05$ & Low density \\
\citet{Li2026} & $z \sim 5$--$7$ & $8.44$\tablefootmark{d} & $0.30 \pm 0.05$ & High density \\
\hline
\end{tabular}
\tablefoot{The relation is parameterized with a normalization $Z_0$, the metallicity at $M_* = 10^{10}\,M_\odot$, and a slope $\gamma$: $12+\log({\rm O/H}) = Z_0 + \gamma\,[\log(M_*/M_\odot) - 10]$.
\tablefoottext{a}{Our five LAEs combined with the galaxies with stellar masses from \citet{hsiao2026} and the auroral-line galaxies of \citet{morishita2024diverse}; galaxies in both samples are counted once.}
\tablefoottext{b}{Smoothly broken power law, $12+\log({\rm O/H}) = Z_0 - (\gamma/\beta)\log[1+(M_*/M_0)^{-\beta}]$, with $Z_0 = 8.793 \pm 0.005$, $\log(M_0/M_\odot) = 10.02 \pm 0.09$, $\gamma = 0.28 \pm 0.02$, and $\beta = 1.2 \pm 0.2$.}
\tablefoottext{c}{Normalization at $\log(M_*/M_\odot) = 8$, as published.}
\tablefoottext{d}{Evaluated at $\log M_* = 10$ from their fits $12+\log({\rm O/H}) = a\log M_* + b$, with $b = 5.04 \pm 0.37$ (low density) and $5.44 \pm 0.40$ (high density); the uncertainty at $\log M_* = 10$ cannot be derived without the covariance between $a$ and $b$, which is not reported.}}
\end{table*}

\begin{figure*}[h!]
    \centering
    \includegraphics[width=0.49\linewidth]{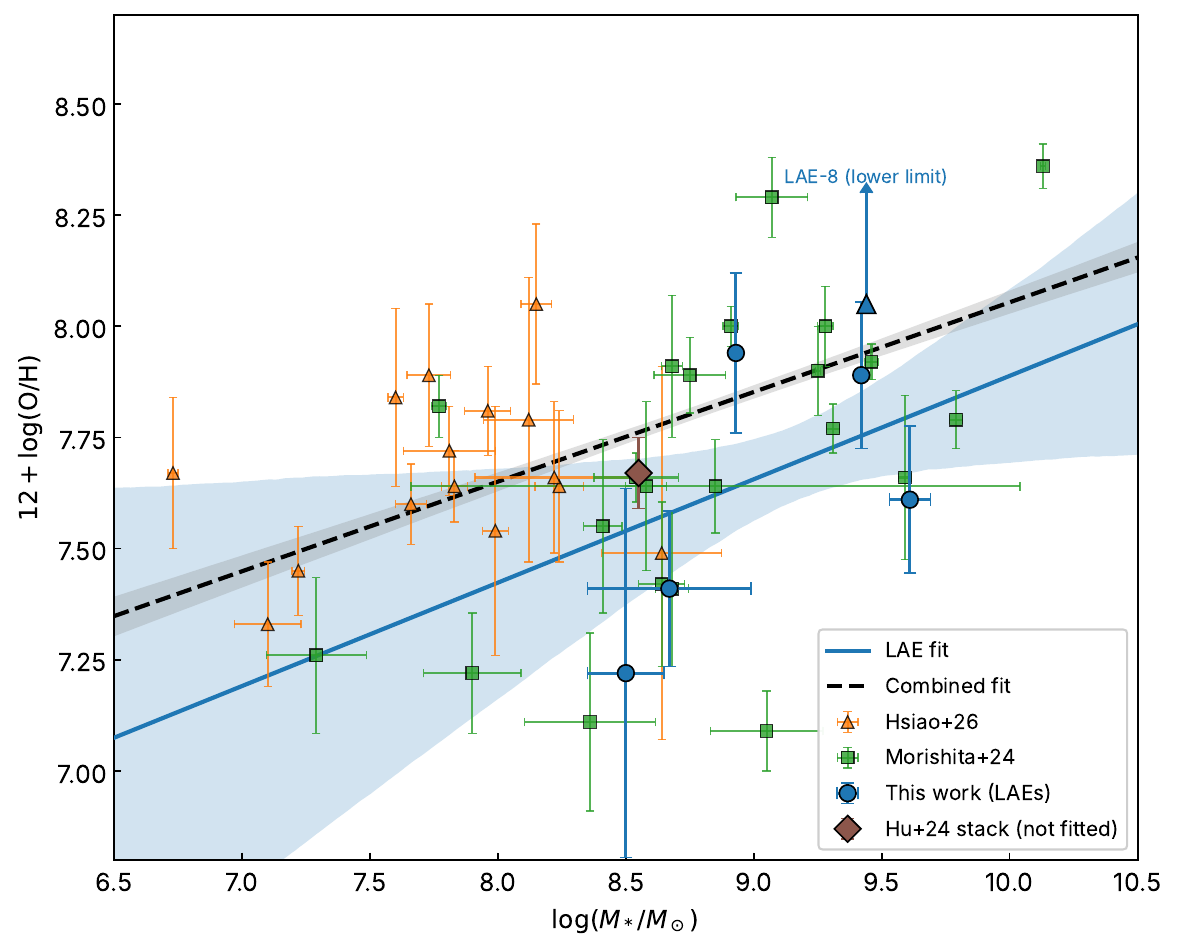}
    \includegraphics[width=0.49\linewidth]{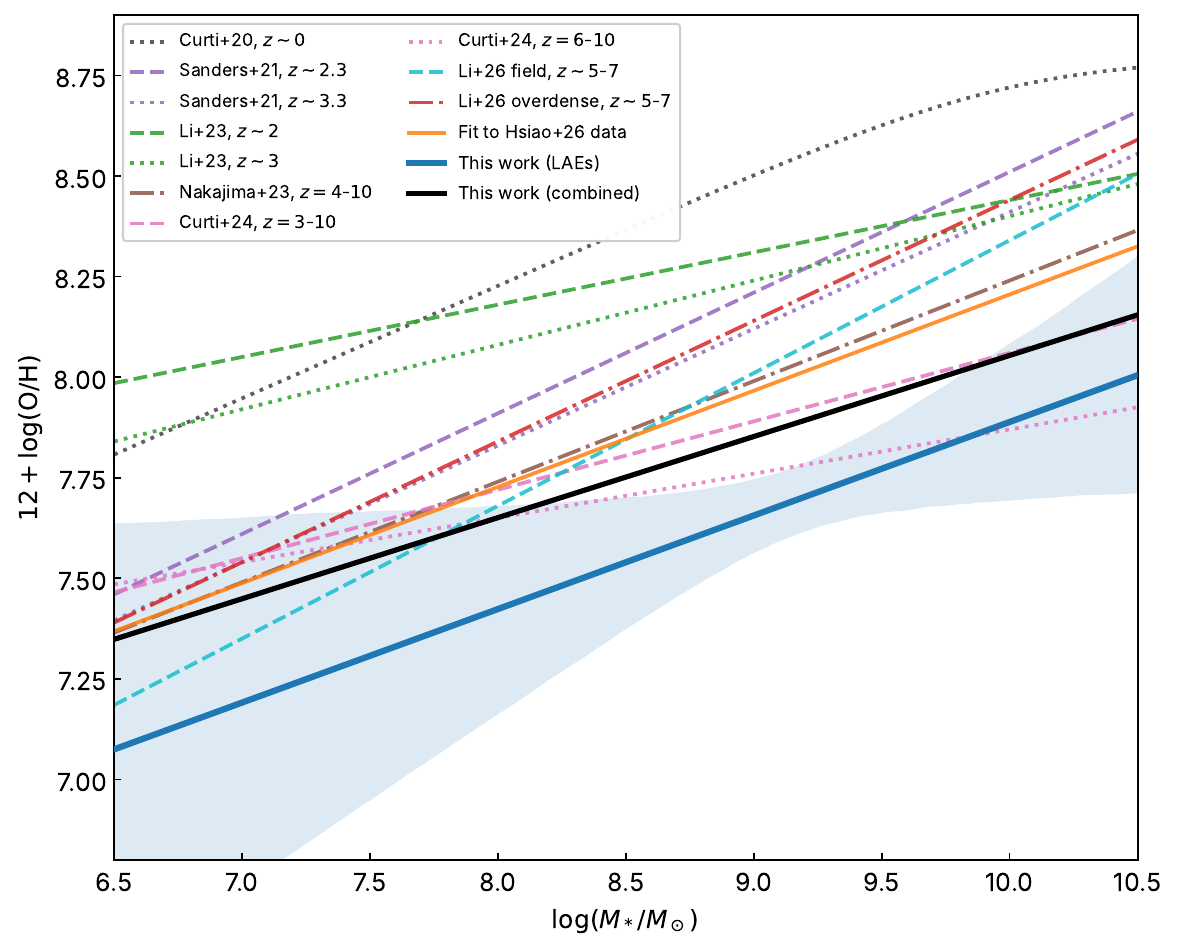}
    \caption{MZR and comparison with literature relations. Left: Our five $z \sim 6.9$ LAEs with auroral-line detections (blue circles), the galaxies with \oauroral detections and stellar masses from \citet{hsiao2026} (orange triangles), and the auroral-line galaxies at $z = 3.2$--$9.4$ from \citet{morishita2024diverse} (green squares); galaxies present in both samples are shown once. LAE-8 is shown as a lower limit (upward triangle) and is excluded from the fits. The brown diamond marks the stack of $5.6<z<9$ galaxies from \citet{Hu2024}, which is not included in the fits. We overlay two linear fits of the form $12+\log(\rm O/H) = Z_0 + \gamma\,(\log M_*/M_\odot - 10)$: one to our LAEs alone (solid blue line; $Z_0=7.888 \pm 0.199$, $\gamma=0.232 \pm 0.214$) and one to all samples combined (dashed black line; $Z_0=8.054 \pm 0.026$, $\gamma=0.202 \pm 0.018$). Shaded bands show the 68\% intervals from 5000 Monte Carlo draws of the fit parameters, including their covariance. Right: Our LAE-only (blue) and combined (black) fits compared with the literature relations of Table~\ref{tab:mzr_params}: the local relation of \citet{curti2020mass}, the $z \sim 2.3$ and 3.3 relations of \citet{Sanders2021}, the $z \sim 2$ and 3 dwarf-galaxy relations of \citet{li2023mass}, the $z = 4$--$10$ relation of \citet{nakajima2023jwst}, the $z = 3$--$10$ and $6$--$10$ relations of \citet{Curti2024}, the field and overdense $z \sim 5$--$7$ relations of \citet{Li2026}, and our fit to the \citet{hsiao2026} sample. All relations are drawn over the full plotted mass range, which extrapolates some of them beyond the stellar masses of their original samples. Our LAE relation lies below the lower-redshift relations and is consistent within its uncertainties with the $z = 6$--$10$ relation of \citet{Curti2024}; its slope agrees within the uncertainties with those of the high-redshift relations.}

    \label{fig:MZR}
\end{figure*}

Stellar masses for our sample were derived from BEAGLE SED fitting. To investigate the evolutionary state of our galaxies, we fitted the MZR using the linear parameterization $12 + \log(\rm O/H) = Z_0 + \gamma\,[\log(M_*/M_\odot) - 10]$, where $Z_0$ is the metallicity at $M_* = 10^{10}\,M_\odot$. The best-fit parameters are summarized in Table \ref{tab:mzr_params}. Figure \ref{fig:MZR} shows our derived relation and relations from the literature.

For our protocluster sample alone, we obtain a slope of $\gamma = 0.232 \pm 0.214$ and a normalization of $Z_0 = 7.888 \pm 0.199$. When the auroral-line galaxies of \citet{hsiao2026} and \citet{morishita2024diverse} ($z = 3.2$--$9.4$) are included in a combined fit, the relation tightens to a slope of $\gamma = 0.202 \pm 0.018$ and a normalization of $Z_0 = 8.054 \pm 0.026$. 

These parameters reveal a distinct chemical offset when compared to established MZRs at lower redshift. At $M_* = 10^{10}\,M_\odot$, the $z \sim 2$--$3$ relations of \citet{Sanders2021} reach $12+\log(\rm O/H) = 8.41$--$8.51$, and the $z = 4$--$10$ relation of \citet{nakajima2023jwst} reaches 8.24, whereas our protocluster LAEs ($Z_0 \approx 7.89$) lie systematically below these relations. This shows that, for a given stellar mass, our sources hold fewer metals in their ISM than galaxies at lower redshift. Our LAEs are, however, consistent within the uncertainties with the $z = 6$--$10$ relation of \citet{Curti2024}, as expected for galaxies at $z \approx 7$; an offset from lower-redshift relations alone therefore does not isolate an environmental effect. We also compare with the lower-stellar-mass sample from \citet{hsiao2026}; their sources lie mostly above our MZR relation but remain consistent with the increasing scatter toward lower masses.

This dilution for a fixed stellar mass may nonetheless reflect the unique environment of the protocluster, as we discuss below. Large reservoirs of pristine gas are expected to be found within the overdensity; as this pristine gas is rapidly accreted onto the galaxies, it fuels enhanced star formation and accelerates mass buildup, while simultaneously diluting the overall metal-to-hydrogen ratio of the ISM.

Crucially, the environmental driver of this dilution is further illuminated when comparing our results to the MZRs of early overdensities presented by \citet{Li2026}. Using JWST strong-line diagnostics for galaxies at $z \sim 5-7$, \citet{Li2026} found that typical galaxies in overdense regions are systematically more metal-enriched ($\sim 0.1-0.2$ dex) than their field counterparts at fixed stellar mass, which is interpreted as being the result of accelerated assembly and efficient metal retention.

Our LAGER-z7OD1 sample stands in stark contrast to this typical environmental enrichment. At a stellar mass of $\log(M_*/M_\odot) \approx 9.1$, the \citet{Li2026} cluster relation predicts a metallicity of $12+\log(\rm O/H) \approx 8.15$. Our LAGER-z7OD1 stacked measurement of $7.76^{+0.06}_{-0.05}$ at a comparable mass lies substantially below both their cluster and field relations. This severe offset can be understood through the combination of our specific selection criteria and the evolutionary phase of the overdensity. Because our sample is strictly composed of LAEs, we are preferentially selecting the most actively star-forming, gas-rich members of the protocluster. These specific regions are likely undergoing an exceptionally violent phase of pristine gas accretion from the IGM. This massive influx of metal-poor gas rapidly fuels star formation and LAE activity while simultaneously overpowering the normal environmental metal retention seen in more average protocluster members, thereby systematically diluting the interstellar media of our observed galaxies.

\subsection{Implications for galaxy evolution in protoclusters}
Our results indicate a complex interplay between rapid mass buildup and ongoing gas accretion within the LAGER-z7OD1 protocluster. This accelerated evolution allows protocluster LAEs to reach higher stellar masses (and therefore higher absolute metallicities) than LAEs in the general field population. At the same time, large reservoirs of low-metallicity gas that fuel star formation---and hence Lyman-$\alpha$ emission---can dilute the ISM, lowering the oxygen-to-hydrogen abundance ratio.

Larger samples with auroral-line detections, particularly in the dense protocluster core and combined with rest-frame infrared photometry, will be needed to disentangle these accretion and feedback cycles and to constrain spatial variations in metallicity across the overdensity. In addition, samples of LAEs spanning a range of environments will be essential to test whether LAEs in dense groups exhibit lower metallicities at fixed stellar mass than their field counterparts.

\section{Conclusions}

We have presented the first direct ($T_e$-based) gas-phase metallicity measurements for a sample of LAEs located inside a known protocluster at $z \approx 7$ (LAGER-z7OD1). Using deep JWST/NIRSpec G395H spectroscopy with \oauroral detections, we derived oxygen abundances for a strictly defined sample of six sources, yielding five secure detections of the [O~{\scriptsize III}]$\lambda$4363 auroral line and one upper limit. 

The metallicities of the secure detections span the range $12 + \log(\rm O/H) \approx 7.2 - 8.0$, while a weighted stack of five sources yields $12 + \log(\rm O/H) = 7.76^{+0.06}_{-0.05}$. All measurements are reported using the observed (uncorrected) line fluxes as a baseline, with the systematic uncertainty on dust extinction derived from the H$\beta$/H$\gamma$ Balmer decrement fully propagated into the final asymmetric error budget via Monte Carlo simulation.

We compare the metallicities of our LAEs with values measured from stacked spectra of LAEs and from photometrically selected galaxies at similar redshifts. Our sources are more metal-rich than the LAE stack and have metallicities closer to those of the photometrically selected population. This implies that, at fixed metallicity, protocluster LAEs are likely to be more massive than their field counterparts.

By combining these direct-method metallicities with stellar masses derived from multiband photometry, we find a complex environmental interplay driving early galaxy evolution. Our stacked protocluster sample exhibits metallicities roughly $0.26$ dex higher than stacks of photometrically selected, general population LAEs at similar redshifts. However, when examining the MZR, our sample yields a slope of $\gamma = 0.232 \pm 0.214$ and a normalization of $Z_0 = 7.888 \pm 0.199$; this relation lies systematically below the established relations at $z \sim 2$--$3$ and $z = 4$--$10$, while remaining consistent with the $z = 6$--$10$ relation of \citet{Curti2024}. The large errors come from our low number statistics and allow for a flat MZR at the 2$\sigma$ level. Such a relation would give a constant metallicity for all mass ranges; however, we see a notable difference between the highest and lowest metallicity values in the sample.

We interpret these combined findings as evidence of an accelerated, yet highly dynamic, evolution unique to the dense protocluster environment. As recently highlighted by \citet{Li2026}, typical protocluster galaxies at $z \sim 5-7$ exhibit environmental metal enrichment due to early assembly. However, the severe metallicity deficit in LAGER-z7OD1 indicates that we are observing a distinct population. Because our selection strictly targets actively star-forming LAEs, we are likely capturing a specific, violent phase of pristine gas accretion. This massive influx of metal-poor gas from the IGM rapidly fuels star formation while overpowering normal environmental metal retention; it thereby systematically dilutes the interstellar media of these specific protocluster members and forces them below the expected MZR. Nevertheless, our sources appear to be more evolved than LAEs not in protocluster environments as shown by the difference with stacks.

Higher number statistics are required to confirm or reject the proposed metal enrichment in protoclusters, either by further observing LAGER-z7OD1 or by observing new dense fields. As LAEs tend to be low-metallicity systems, detection and follow-up of other kinds of galaxies is necessary to study the structure in detail. Such objects could be submillimeter galaxies (SMGs) or Lyman-break galaxies (LBGs), which could be detected with the Atacama Large Millimeter/submillimeter Array (ALMA). In particular, detecting the core of the protocluster will allow us to study spatial dependences more precisely. Lastly, building well-defined samples of sources across a range of cosmic environments will be key in illuminating the effects of field conditions on galaxy evolution.

\begin{acknowledgements}
CMS acknowledges support from Beca ANID folio 21211528 and support from CATA. This paper includes data gathered with JWST. CMS and LFB acknowledge support from ANID BASAL project FB210003 and FONDECYT grant 1230231, and the China-Chile Joint Research Fund (CCJRF No. 1906). JXW acknowledges support from the science research grant from the China Manned Space Project with No. CMS-CSST-2021-A07. J.R.W. acknowledges that support for this work was provided by The Brinson Foundation through a Brinson Prize Fellowship grant.
Support for program JWST-GO-01635 was provided by NASA through a grant from the Space Telescope Science Institute, which is operated by the Association of Universities for Research in Astronomy, Inc., under NASA contract NAS 5-03127. 
We thank the anonymous referee for meaningful feedback and commentary on the manuscript for this paper.
\end{acknowledgements}
\bibliographystyle{aa}
\bibliography{aa58553-25}
\end{document}